\documentclass[reprint, twocolumn,
showpacs,
amsmath,amssymb,aps,prl, superscriptaddress]{revtex4}
\usepackage{graphicx}
\usepackage{dcolumn}
\usepackage{bm}

\newcommand{\vrec}{\ensuremath{v_\mathrm{rec}}}

\begin{document}

\title{Sub-Poissonian number differences in four-wave mixing of matter waves}

\author{J.-C.~Jaskula} 
\author{M.~Bonneau}
\author{G.~B.~Partridge}
\affiliation{Laboratoire Charles Fabry de l'Institut d'Optique, CNRS, Univ Paris-Sud, Campus Polytechnique RD128 91127 Palaiseau France}

\author{V.~Krachmalnicoff}
\altaffiliation{Present address: Institut Langevin, ESPCI Paris Tech, CNRS, Paris France}
\affiliation{Laboratoire Charles Fabry de l'Institut d'Optique, CNRS, Univ Paris-Sud, Campus Polytechnique RD128 91127 Palaiseau France}

\author{P.~Deuar}
\affiliation{Institute of Physics, Polish Academy of Sciences, Al. Lotnik\'{o}w 32/46, 02-668 Warsaw, Poland}

\author{K.~V.~Kheruntsyan}
\affiliation{ARC Centre of Excellence for Quantum-Atom Optics, School of Mathematics and Physics,University of Queensland, Brisbane, Queensland 4072, Australia}

\author{A.~Aspect}
\author{D.~Boiron}
\author{C.~I.~Westbrook}
\affiliation{Laboratoire Charles Fabry de l'Institut d'Optique, CNRS, Univ Paris-Sud, Campus Polytechnique RD128 91127 Palaiseau France}

\date{\today}

\begin{abstract}
We demonstrate sub-Poissonian number differences in four-wave mixing of Bose-Einstein condensates of metastable helium. The collision between two Bose-Einstein condensates produces a scattering halo populated by pairs of atoms of opposing velocities, which we divide into several symmetric zones. We show that the atom number difference for opposing zones has sub-Poissonian noise fluctuations whereas that of nonopposing zones is well described by shot noise. The atom pairs produced in a dual number state are well adapted to sub shot-noise interferometry and studies of Einstein-Podolsky-Rosen-type nonlocality tests.
\end{abstract}

\maketitle

The creation of squeezed states of the electromagnetic field has been a major preoccupation of quantum optics for several decades~\cite{bachor:04}. 
Such states are not only inherently fascinating, but they also have the potential to improve sensitivity in interferometers~\cite{bachor:04}, going beyond the ``shot noise" or standard quantum limit. 
In the field of atom optics, workers are beginning to use the intrinsic non-linearities 
present in a matter wave field to produce non-classical states, 
especially squeezed states \cite{chuu:05,Whitlock:10,Itah:10,esteve:08,riedel:10,maussang:10}. Indeed, 
an atom interferometer using squeezed inputs  was recently demonstrated~\cite{Gross:10}. In our case, we produce dual number states in four-wave mixing of Bose-Einstein condensates (BECs). These states form the basis of a very different proposal for  atom interferometry beyond the standard quantum limit~\cite{bouyer:97,dunningham:02,campos:03}. 
Squeezing of atom samples may prove even more important than squeezing of light
because the number of available atoms is often limited and therefore
surpassing the standard quantum limit can be the only way to increase the signal-to-noise ratio and improve performance. 
In interferometry proposals relying on dual number states, the observable corresponding 
to the relative phase is completely undetermined. 
Paradoxically, after passing through a beam splitter, the phase difference is no longer 
undetermined, 
but is peaked with a dispersion below the shot noise~\cite{dunningham:02,campos:03}.
It has been argued that such states can be more robust to loss processes than
maximally entangled states~\cite{dunningham:02}. 
The pairs we produce should also be entangled in a sense analogous
to \cite{rarity:90}. 
A potentially interesting feature of our stuation is that the pairs
have large spatial separations (several cm here) and 
are thus well suited to investigations of 
(non-local) EPR entanglement \cite{Reid:09} and Bell's inequalities using atoms.

Correlated photon pairs can be generated using optical processes such as 
four-wave mixing \cite{Boyer:08} or parametric down conversion \cite{burnham:70}.
The matter wave analogs of these processes have recently been 
demonstrated \cite{greiner:05, Perrin:2007}. 
The spontaneous four-wave mixing process \cite{Perrin:2007}, which we use here,
simply corresponds to the collision of two Bose-Einstein condensates during which
binary collisions produce scattered pairs of atoms with correlated momenta. 
Correlations however, do not guarantee relative number squeezing 
(see Ref.~\cite{buchmann:10} for an example) 
nor entanglement. 
The success of proposals such as those of Refs.~\cite{bouyer:97,dunningham:02,campos:03} 
will likely be determined by the degree of squeezing. 
Thus, with a view towards using such correlated states in interferometry, 
it is important to verify that these processes do indeed produce squeezing. 
In this letter, we demonstrate and quantify sub-Poissonian number differences produced 
in this process. 
Although the observation is not strictly sufficient to demonstrate squeezing
in the sense of measuring fluctuations in two conjugate variables, 
we will often use the term squeezing below because the situation
is a close atomic analog to experiments such as  
Ref.~\cite{Heidmann:1987} in which relative intensity 
squeezing was observed in the generation of twin light beams created by parametric down conversion.

\begin{figure}[t]
\begin{center}
\includegraphics[width=\linewidth]{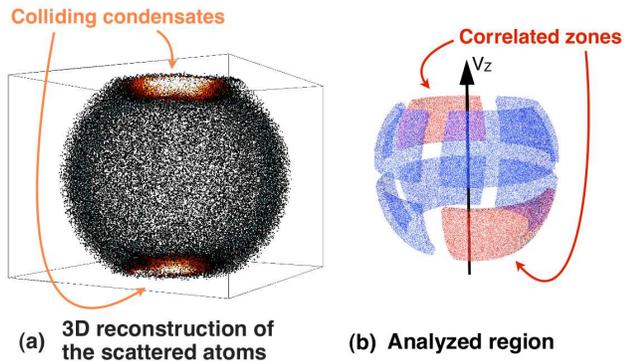}
\caption{\label{fig:Sphere}(Color online) View of the 
halo after the collision of two BECs and a subsequent ballistic expansion.
(a) The experimental data plotted in momentum space, with each dot corresponding to a detected atom. 
Atoms on the collision halo are black, while
the colliding, pancake shaped BECs at the top and the bottom of the halo are orange/yellow.
The collision axis $v_z$ and the optical trap axis are both almost vertical.
(b) Schematic view of the analyzed part ($|v_z| <  0.5 \, \vrec $)
of the
collision halo. 
Here we use $N_Z=8$ zones that are separated from each other for better visualization. 
An example of two correlated zones is shown (red,arrow). 
The number difference between these two zones shows sub shot-noise fluctuations. 
}
\end{center}
\end{figure}

We use metastable helium atoms which are detected by a micro-channel plate detector 
with a delay line anode \cite{schellekens:05}.
The detector allows three dimensional reconstruction of the momentum of each atom.
Atoms in the  $2^3S_1,\, m_x=1$ state are evaporatively cooled in a 
vertically elongated optical trap to produce a BEC
with about $10^5$ atoms and no discernible thermal component~\cite{Partridge:2010}. The use of an optical trap has resulted in substantially better shot to shot reproducibility
than its magnetic antecedent \cite{Perrin:2007}.
The atomic angular momentum, which is due entirely to the electron spin, 
is defined relative to a 4 G magnetic holding field in the $x$-direction 
(orthogonal to the optical trap axis).
After cooling, the atomic spin is rotated away from the axis of the holding field
by $\pi /2$ using a 2 ms rf sweep \cite{Partridge:2010}. 
The laser trap is then switched off and 
$1~\mu$s later the condensate is split by applying counter-propagating
laser beams for $2.5~\mu$s. These beams are blue detuned from the $2 ^3P_0$ state by 600~MHz, inclined at a $7^\circ$ angle to the vertical axis and linearly polarized along the quantization axis. About one third of the atoms are diffracted into each of two momentum classes traveling at $\pm 2\,\vrec$, where $\vrec = 9.2$~cm/s is the recoil velocity. Most of the rest remain at zero velocity. 
Binary collisions take place between atoms of all three velocity classes producing three collision halos with center of mass velocities $\pm \vrec$ and zero. Since the atomic spin is orthogonal to the local field, 50\% of the atoms are in the
$m_x=0$ state with respect to the magnetic field axis~\cite{Partridge:2010}, and these atoms fall to
the detector, unperturbed by magnetic field gradients. 
The trajectories of atoms in the $m_x=\pm 1$ states are perturbed by residual field 
gradients and we therefore apply an additional gradient that causes these atoms to 
miss the detector entirely. The analysis is only focused on the collision halo centered at $+\vrec$~(see Fig.~\ref{fig:Sphere}a). 

The collision halo centered at $v = 0$ has a radius $2\,\vrec$ and is too large to be entirely captured by the detector
while the two halos centered at $\pm \vrec$, with radii $\vrec$, are entirely detected. 
In addition to binary scattering events, these two latter halos can be populated by spontaneous photon scattering whenever an atom at $v = 0$ scatters a photon from one of the diffraction laser beams. The diffraction efficiency depends on 
the product $I_1 I_2$ of the two laser intensities, 
while the spontaneous scattering into a given halo
depends on only one of these intensities. So to reduce this effect we introduce an intensity imbalance in the two laser beams such that the halo centered at $+\vrec$ is populated by the weaker beam and contains fewer such optically scattered atoms. 
\begin{figure}[t]
\begin{center}
\includegraphics[width=\linewidth]{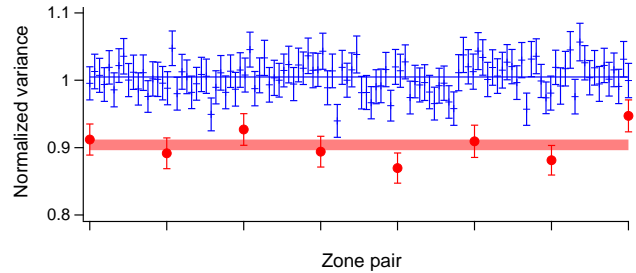}
\caption{(Color online) Variance of all possible pairs of zones for the halo cut into 16 zones 
and summing $N_s=3600$ shots. The normalized variance is  $V_{i,j}$ and the error bars reflect its standard deviation $\delta V_{i,j}$ with $\delta V^2_{i,j}=\frac{1}{N_s}\frac{\langle (N_i-N_j)^4\rangle-\langle (N_i-N_j)^2\rangle^2}{\langle N_i+N_j\rangle}$. 
Circles correspond to the 8 correlated zones and crosses to the 112 uncorrelated ones. The two horizontal lines correspond to the mean of each data set with a thickness given by twice the standard deviation of the mean, considering each pair of zones as independent. 
}

\label{fig:SqueezedParts}
\end{center}
\end{figure}

If squeezing is present, we expect a sub-shot noise variance in the number difference of any two 
diametrically opposed volumes in the scattering halo \cite{savage:07}. 
For any other pair of volumes, we expect a variance corresponding to shot noise. 
We define the halo as a spherical shell of radius $\vrec$ and 
thickness $\pm 0.15\, \vrec$.
The results are only weakly sensitive to this thickness
but as defined, it includes about 95\% of the scattered atoms.
We remove the areas on the halo
containing the scattered BEC's. 
The excised regions correspond to 
vertical velocities $|v_z| >  0.5\, \vrec $.
We divide the remainder of the halo in half at the equator and 
then make $p$ vertical cuts along the meridians,
dividing the halo into $N_Z=4p$ equal zones as shown in Fig.~\ref{fig:Sphere}b for $p=2$.
We define a normalized number difference variance for zones $i$ and $j$:
\begin{equation}
V_{i,j}=\frac{\langle(N_i - N_j)^2\rangle - \langle N_i  - N_j\rangle^2}{\langle N_i \rangle+\langle N_j\rangle}
\label{eq:numdiff}
\end{equation}
The brackets $\langle ...\rangle$ denote the average over the 3600 shots, and
$N_i$ refers to the number of atoms detected in the {\it i}-th zone on a single shot.
On average, we detect 150 atoms per shot on the whole analyzed region.
If the zones $i$ and $j$ are uncorrelated, the normalized variance should be unity.
 Figure~\ref{fig:SqueezedParts} shows the measured variances of all possible pairs of zones when the halo is cut into 16 zones. 
The eight pairs of correlated zones indeed show 
sub-Poissonian number differences ($V<1$) and the 112 pairs of uncorrelated zones do not.

Perfectly correlated pairs and perfect detection would result in a zero variance.
This however is almost unattainable in practice because of various imperfections, 
the most significant of which is 
the non-unit quantum efficiency $\eta$ of our 
detector. 
The effect of the efficiency alone leads to a variance $V=1-\eta$ of the correlated zones,
and therefore we can immediately deduce
a lower limit of 10\% on the quantum efficiency, in agreement with estimates
we have made in the past \cite{Jeltes:07}. 

\begin{figure}[t]
\begin{center}
\includegraphics[width=\linewidth]{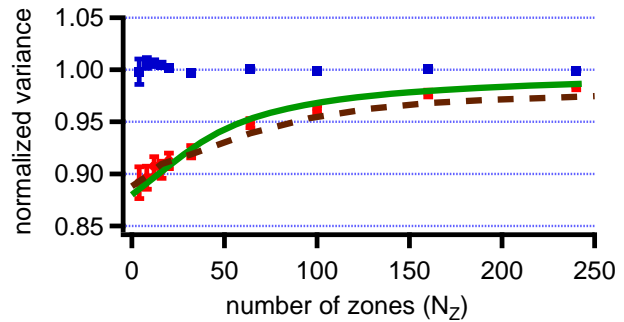}
\caption{(Color online) Observed variance, as a function of the number of zones into which we cut
the halo. Red circles: average over all correlated zones, 
blue squares:  average over all uncorrelated zones.
Error bars show the standard deviation of the mean 
of the variances for a given $N_Z$.
The solid curve is the prediction of the simple model discussed in the text. 
The dashed curve 
results from the stochastic Bogoliubov simulation.
Both models assume a 12\% quantum efficiency.
}
\label{fig:SqueezingVsCut}
\end{center}
\end{figure}

A second, less severe but intrinsic imperfection comes about because the momenta of
the correlated atoms are not exactly equal and opposite, but have a width 
determined by the momentum spread within the initial condensates, 
as confirmed by the finite width of the two-body correlation 
function in momentum space \cite{Perrin:2007}.
Thus it is possible for the two atoms of a correlated pair to end up in
zones that are not diametrically opposed. 
We can study this effect by observing how the 
variance changes as we change
the number of zones $N_Z$ (Fig.~\ref{fig:SqueezingVsCut}). 
The smaller the zones, 
the more likely
that an atom will miss the zone diametrically opposed to that of its partner. 

Since we have measured the correlation function for back-to-back momenta, 
we can model  the trend seen in 
Fig.~\ref{fig:SqueezingVsCut}.
The back-to-back correlation function was measured to have rms widths of 
$0.17\,\vrec$ in the radial ($x$ and $y$) directions, 
and $0.02\,\vrec$ in the axial ($z$) direction.
Neglecting the much smaller axial correlation width, 
we estimate the probability $P(N_Z)$ that, given an atom hitting one zone, 
its partner will hit the diametrically opposite one. 
This probability decreases as $N_Z$ increases, 
and, taking both quantum efficiency and the geometrical hit probability into account, we expect $V=1-\eta P$. 
The function $V(N_Z)$ is plotted as the solid line in Fig.~\ref{fig:SqueezingVsCut}.
The approximate agreement of this simple model with the data
leads us to conclude that the above two loss mechanisms
account very well for the observed variance. 
We also get a slightly better lower limit on the quantum efficiency, 
$\eta >12\%$.

The situation was also analyzed using a stochastic Bogoliubov simulation
as in Ref.~\cite{Krachmalnicoff:10}.
The result for the variance is shown as the dashed curve in Fig.~\ref{fig:SqueezingVsCut}.
The curve is plotted assuming a detector quantum efficiency of 12\% as in the
simpler model.
The simulation shows the observed trend, but agrees less well with the data
than the simple model.  
The discrepancy arises because the simulation predicts a narrower back-to-back correlation 
function than was observed in the data, 
thus resulting in a slower approach to unity for the variance.
A finite temperature effect may be at the origin of the difference since the simulation 
assumes zero temperature.
The simulation also neglected mean field repulsion of different spin components, 
so that such effects could also be responsible. 
The calculation nevertheless confirms the idea that the lack of perfect
correlation in momentum determines most of the variation 
seen in Fig.~\ref{fig:SqueezingVsCut}.

Other known imperfections include the possible contamination of the sphere
by atom pairs with one atom in the $m=0$ state
and another in the $m=1$ state. 
These pairs contribute a single detected atom without a partner to the halo.
We have no independent experimental estimate of the number of such collisions
but they could account for as much as one half of the observed atoms on the halo. 
Their presence would mimic a loss in detector quantum 
efficiency and thus raise our lower limit on $\eta$.
Spontaneous emission processes  act in the
same way, but independent measurements indicate that  such processes contribute only about 1.5\% of the detected atoms on the analyzed halo. As discussed above, the halo centered at $-\vrec$ was more affected by spontaneous emission, though squeezing is still also observed, albeit to a lesser degree. 
While one might hope to improve the quantum efficiency of the detector, 
or suppress unwanted scattering events, 
the stochastic Bogoliubov simulation with $\eta=1$ 
predicts a limiting variance
$V\approx 0.1$ for a small number of zones. 
Thus, correcting for the quantum efficiency, 
the intrinsic squeezing appears to be at most $-10$~dB.

Relative number squeezing is also related to the violation of
a classical Cauchy-Schwarz inequality~\cite{walls:08, perrin:08},
\begin{equation}
\left\langle N_i N_j \right\rangle \le
\sqrt{\left\langle N_i^2\right\rangle \left\langle N_j^2 \right\rangle}\:,
\label{eq:cs}
\end{equation}
relating the count rates in two correlated zones $i$ and $j$.  
For equal count rates in the two zones, relative number squeezing is strictly 
equivalent to the violation of the inequality \eqref{eq:cs}.
In our experiment 
the average count rates are not exactly equal, 
in which case squeezing and Cauchy-Schwarz violation are not equivalent \cite{marino:08}.
Never the less, we do  observe a violation of the inequality  \eqref{eq:cs}.
More sophisticated inequalities can also be invoked
and will be studied in future work.

For purposes of interferometry,
one would like to increase count rates and the number of atoms per mode. This could be achieved in a four-wave mixing experiment inside an optical lattice 
to modify the dispersion relations of the atoms so as to populate a single pair of modes~\cite{Hilligsoe:05,campbell:06,gemelke:05}. 
Such well defined twin atom beams would permit the realization
of experiments 
such has the celebrated experiment of Hong, Ou and Mandel \cite{hong:87}, or the realization of an interferometer in the spirit of \cite{bouyer:97,dunningham:02,campos:03}.
Even more ambitious would be the demonstration of entanglement of the pairs
by making Bell-type measurements
of the well separated neutral 
atoms, in analogy with the measurement made in 
Ref.~\cite{rarity:90} using photons.

This work was supported by the French ANR, the IFRAF Institute, and
the Euroquam Project CIGMA.  
G. P. is supported by a European Union Marie Curie IIF Fellowship.
P. D. acknowledges the EU contract No. PERG06-GA-2009-256291.
K. K. acknowledges support from the Australian Research Council.

%

\end{document}